\documentclass[aps, twocolumn,superscriptaddress,preprintnumbers,amsmath,amssymb]{revtex4}

\usepackage{graphicx}
\usepackage{latexsym}
\usepackage{dcolumn}
\usepackage{bm}
\usepackage{float}
\usepackage{graphicx,here}
\usepackage{color}
\usepackage{soul}
\usepackage{siunitx}

\begin{document}

\title{Electronic structure of a 3x3-ordered silicon layer on Al(111)}

\newcommand{\AffISSP}{\affiliation{Institute for Solid State Physics (ISSP), The University of Tokyo, Kashiwa, Chiba 277-8581, Japan}}
\newcommand{\AffASRC}{\affiliation{Advanced Science Research Center, Japan Atomic Energy Agency,
2-4 Shirakata, Tokai, Naka, Ibaraki, 319-1195, Japan}}
\newcommand{\AffNanoparis}{\affiliation{Institut des NanoSciences de Paris, Sorbonne Université, F-70005 Paris, France}}
\newcommand{\AffMineral}{\affiliation{Institut de Minéralogie, de Physique des Matériaux et de Cosmochimie, Sorbonne Université, F-70005 Paris, France}}
\newcommand{\AffSorbonne}{\affiliation{Sorbonne Université, Institut des NanoSciences de Paris (INSP), F-75005 Paris, France}}
\newcommand{\AffISMCNR}{\affiliation{Istituto di struttura della Materia, Consiglio Nazionale delle Ricerche,
Strada Statale 14 km 163.5, Trieste, I-34149, Italy}}
\newcommand{\AffIMSS}{\affiliation{Institute of Materials Structure Science, High Energy Accelerator Research Organization (KEK), Tsukuba, Ibaraki 305-0801, Japan}}
\newcommand{\AffIMRAM}{\affiliation{Institute of Multidisciplinary Research for Advanced Materials, Tohoku University, Aoba-ku, Sendai 980-8577, Japan}}
\newcommand{\AffNTHU}{\affiliation{Department of Physics, National Tsing Hua University, Hsinchu 30013, Taiwan}}
\newcommand{\AffNCTS}{\affiliation{Physics Division, National Center for Theoretical Sciences, Hsinchu 30013, Taiwan}}
\newcommand{\AffSinica}{\affiliation{Institute of Physics, Academia Sinica, Taipei 11529, Taiwan}}
\newcommand{\AffLASTI}{\affiliation{Laboratory of Advanced Science and Technology for Industry, University of Hyogo,
3-1-2 Koto, Kamigori-cho, Ako-gun, 678-1205, Hyogo}}

\author{Yusuke~Sato}\AffISSP
\author{Yuki~Fukaya}\AffASRC
\author{Mathis~Cameau}\AffNanoparis\AffMineral
\author{Asish K.~Kundu}\AffISMCNR
\author{Daisuke~Shiga}\AffIMSS\AffIMRAM
\author{Ryu~Yukawa}\AffIMSS
\author{Koji~Horiba}\AffIMSS
\author{Chin-Hsuan~Chen}\AffNTHU
\author{Angus~Huang}\AffNTHU
\author{Horng-Tay~Jeng}\AffNTHU\AffNCTS\AffSinica
\author{Taisuke Ozaki}\AffISSP
\author{Hiroshi~Kumigashira}\AffIMSS\AffIMRAM
\author{Masahito~Niibe}\AffISSP\AffLASTI
\author{Iwao~Matsuda}\AffISSP

\date{\today}

\begin{abstract}
    Electronic structure of the 3x3 ordered-phase of a silicon (Si) layer on Al(111) has been studied by angle resolved photoemission spectroscopy (ARPES) technique using synchrotron radiation and modeled by a trial atomic model.
    A closed Fermi surface originating from 
    liniarly dispersing band is identified. A band structure calculation of a trial atomic model of the honeycomb silicene on Al(111) implies that the metallic band originates from the Al-Si hybrid state that has the Dirac cone-like dispersion curves. The Si layer on Al(111) can be a model system of Xene to realize the massless electronic system through the overlayer-substrate interaction.
\end{abstract}

\maketitle

\section{Introduction}

Elemental atomic layers have attracted academic interest in their fundamental physics and they have also found significance in technological applications such as next generation devices\cite{MatsudaTextbook}. A well-known example is the two-dimensional (2D) honeycomb lattice of carbon, graphene\cite{Neto}. In group-IV element, there have been also reports on silicene (Si), germanene (Ge) and stanene (Sn)\cite{Matthes,Lelay,Takamura,Takagi,FukayaSi,Davila,Tang,FukayaGe}. These monatomic layers, termed Xenes, have been widely studied because they host Dirac cones at the K (K') points of the Brillouin zone (BZ), which enables exotic quantum devices\cite{Matthes}. However, the existence of such Dirac bands in silicene has been controversial because the layers have been grown on crystal surfaces and the significance of the contribution of the external substrate effect has been debated by researchers\cite{Paolo1,Paolo2,Feng1,Feng2,SiCal1,SiCal2}. Using angle-resolved photoemission spectroscopy (ARPES), Dirac cones of the silicene layer on Ag(111) were observed along zone-boundaries in the surface BZ of Ag(111), not at the K (K') points of the silicene BZ\cite{Paolo1,Feng2,SiCal1,SiCal2}. The electronic structure is described theoretically in terms of the superstructure-induced splitting of Dirac cones by the substrate-overlayer interaction\cite{Feng2}. This result reveals the nature of heterojunctions at the electric contacts between silicene and electrode in, for example, a transistor\cite{Tao} and also paves the ways for manipulating silicene's electronic properties in the external environment. 

Silicon deposition has been achieved on other metal substrates\cite{Feng2017,Paolo1,Paolo2,Feng1,Feng2,Lelay,Takagi} and a 3x3-ordered phase is formed on Al(111)\cite{Jona,Munoz1,Munoz2}. The surface has been known to consist of a silicon overlayer and, thus, it is of interest to investigate the surface electronic structure.

In the present research, we studied the electronic states of a silicon layer on Al(111) that forms 3x3 surface periodicity by ARPES using synchrotron radiation. Electronic bands of the overlayer are identified through a comparison with those of the clean Al(111) surface\cite{Kevan}. Our results show, the overlayer is metallic and the metallic band has 
linear dispersion curves. Band calculation, based on density functional theory (DFT), is conducted on a trial atomic model of the honeycomb silicene on Al(111). Similarities between the theoretical and experimental band dispersions indicate that the electronic states of the silicon layer are
likely originating from the Al-Si hybridization, which indicates indispensable substrate-overlayer interaction.   

\begin{figure}
    \includegraphics[width=.8\linewidth,keepaspectratio]{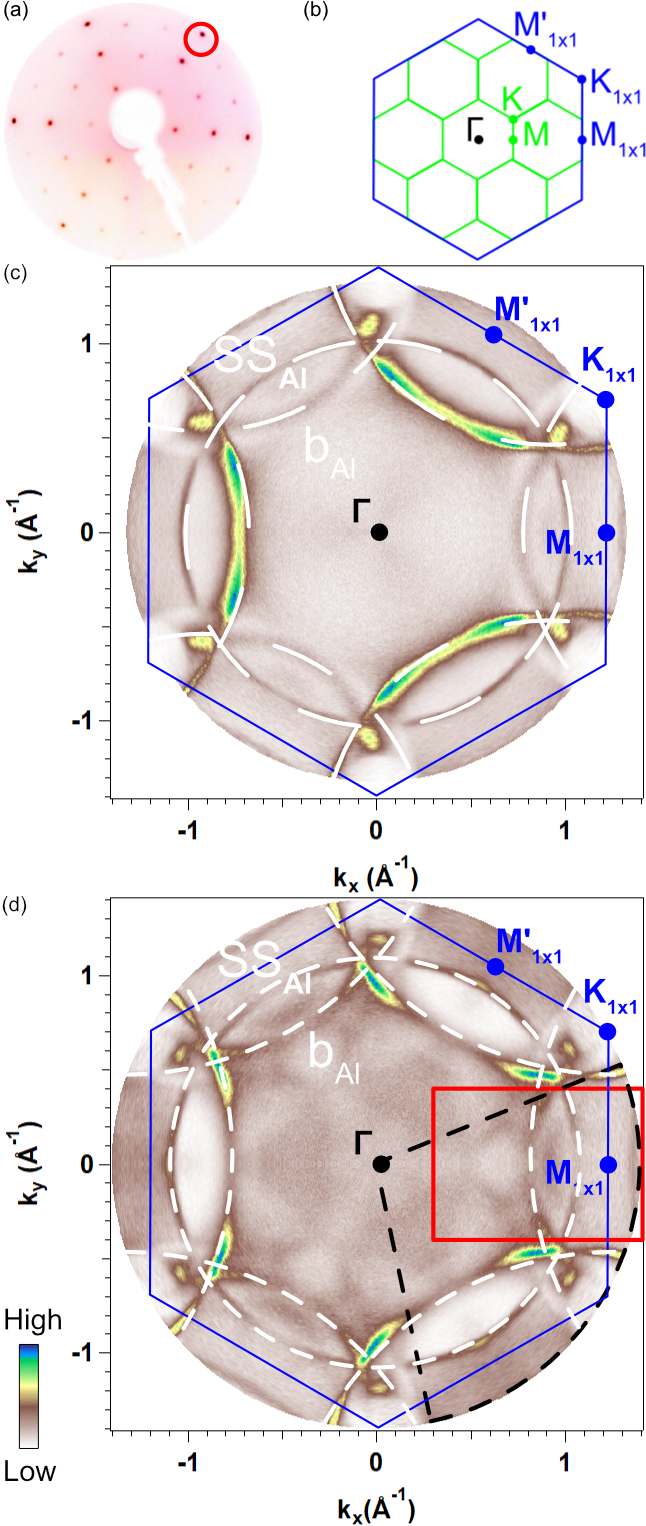}
    \caption{\label{fig1}(a) A LEED pattern at electron energy of 88.2 eV of 3x3-Si/Al(111). Red circle indicates one of the Al-1x1 LEED spot. (b) A schematic diagram of surface Brillouin zone (SBZ). The blue and green hexagons correspond to the Al(111)-1x1 and Al-3x3 SBZ's, respectively. (c) Fermi surface map of 3x3-Si/Al(111) taken at $h\nu=\textrm{45 eV}$ and at 20 K.
    (d) Fermi surface map of 3x3-Si/Al(111) taken at $h\nu=\textrm{47 eV}$ and at 20 K.
    The overall mapping is obtained by the symmetric folding of the region surrounded by black broken curves.
    Fermi surfaces of the clean Al(111) substrate is traced with white broken curves. 
    $\rm SS_{Al}$ and $\rm b_{Al}$ correspond to the Fermi surface originating from surface-resonant/state and bulk bands of Al(111). 
    The red square region is focused in Fig. \ref{fig4}(a).}
\end{figure}

\section{Experimental}
A clean Al(111) surface was obtained by cycles of Ar$^{+}$ sputtering and annealing, followed by confirmation of crystallinity by observing a low-energy electron diffraction (LEED) pattern and of cleanliness by measuring X-ray photoemission spectroscopy (XPS). Subsequently, Si was deposited on Al(111) at 350 K, which showed the 3x3 LEED pattern (Fig. 1(a)). A schematic drawing of the 1x1 surface Brillouin Zone (SBZ) and 3x3 SBZ's is depicted in Fig. 1(b). From the XPS measurement of Al 2$p$ and Si 2$p$ core levels, silicon coverage of the surface overlayer is evaluated about 0.4 ML, which is consistent with previous reports on the 3x3-Si ordered phase on Al(111)\cite{Munoz2}.

ARPES measurements of the 3x3-Si/Al(111) and clean Al(111) surfaces were performed at the VUV-Photoemission beamline (Elettra, Trieste) and at BL-2A (Photon Factory, Japan). The photoemission data were collected using multichannel detection (polar angle) combined with sample rotations (azimuthal or tilt angle) 
and converted into energy ($\rm E$)-momentum space ($\rm k_x$,$\rm k_y$), where $\rm E$ is the kinetic energy and $\rm k_x$,$\rm k_y$ are the in-plane components of the wavenumber (wavevector) parallel to the surface.  

First-principles calculations were performed using the projected augmented wave method (PAW) \cite{Blochl1994,Kresse1999} as implemented in the Vienna Ab initio Simulation Package \cite{Kresse1993,Kresse1994,Kresse1996,Kresse1996a} based on density functional theory with the Perdew-Burke-Ernzerhof (PBE) type of generalized gradient approximation (GGA). The 12x12x1 Monkhorst-Pack k mesh and a cutoff energy of 300 eV are used in the self-consistent field calculations. The geometries of all the systems are optimized with the total energies converged within $10^{-4}$ eV. 
To study the substrate effect in the Si/Al(111) systems, we adopted a trial honeycomb model by geometrically optimizing a 2x2-honeycomb silicene lattice on top of a 7-layer Al(111) 3x3 substrate. As a result of optimization, honeycomb lattice is slightly distorted. The supercell lattice constant is 8.59 {\AA} and interlayer distance is 2.25 {\AA} between the silicene layer and the top Al layer. Bond lengths of the two Si-Si bond in the optimized structure are 2.473 {\AA} and 2.476 {\AA}.
To compare with the experiment Fermi surface, we unfold the band structures of the Silicene 2x2/Al(111) 3x3 supercell Brillouin zone (BZ) to the BZ of the Al(111) 1x1 unit cell by using the BANDUP package\cite{Medeiros}. The Fermi surfaces are derived by tiling the zero energy contour of the unfolded band structure.

\begin{figure}
    \includegraphics[width=.8\linewidth,keepaspectratio]{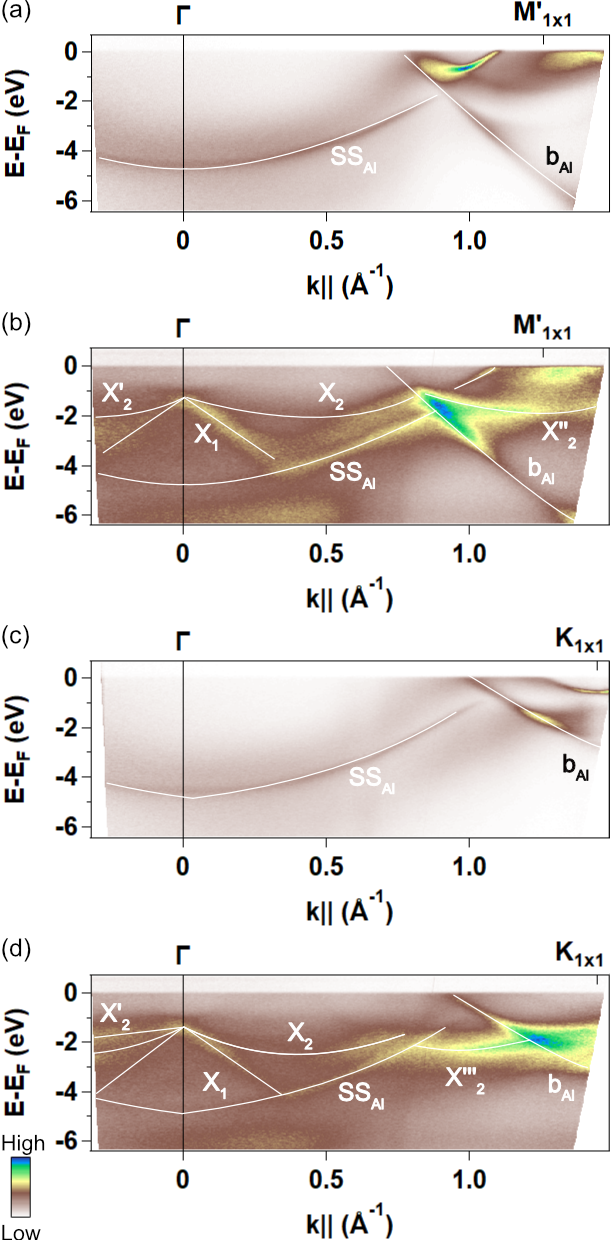}
    \caption{\label{fig2} Band dispersion plots of (a,c) clean Al(111) and (b,d) 3x3-Si/Al(111) along (a,b) $\Gamma$-$M'_{1x1}$ and (c,d)$\Gamma$-$K_{1x1}$ directions taken at $h\nu = \textrm{50 eV}$ and at room temperature. White lines are schematic drawings of the observed band dispersion curves.}
\end{figure}
      
\section{Results and Discussion}
Figure \ref{fig1}(d) displays the photoemission map at the Fermi level ($\rm E_F$) of the 3x3-Si layer on Al(111) with the 1x1 SBZ and symbols of the symmetry points. The intensities of the contours are modulated to give only a threefold symmetry due to the matrix element that is sensitive the local bonding configuration in the material\cite{substrate1,substrate2}. Fermi contours of Al bulk bands, $\rm b_{Al}$, and a surface-resonant/state band, $\rm SS_{Al}$, of Al(111) surface are observed, as previously by ARPES\cite{Kevan}. The clear appearance of these states indicates that the Al surface resonance/state is likely survived after the Si deposition. It leaves a possibility that the Al(111) surface state, generated in the inverted bulk band-gap, is topologically protected as in cases of the noble metal surface-state\cite{Yan}. 
In Fig. \ref{fig1}(d), the intensity distributions, except from $\rm b_{Al}$ and $\rm SS_{Al}$, likely originate from the 3x3-Si/Al(111) system because they are not produced by 3x3 umklapp scattering in either the $\rm b_{Al}$ or $\rm SS_{Al}$ bands. Figure \ref{fig2} shows energy-wavenumber diagrams along symmetry axes of the $\Gamma$-$M'_{1x1}$ and $\Gamma$-$K_{1x1}$ directions, which is used to trace the dispersion curves of the observed features. 
Two types of bands other than $\rm b_{Al}$ and $\rm SS_{Al}$ can be identified, those disperse from the $\Gamma$ point with small ($X_{1}$) and large ($X_{2},X'_{2}$) effective mass. The $X"_{2}$ and $X"'_{2}$ bands, observed at large wavenumbers, likely share curves with the $X_{2}$ ($X'_{2}$) band. 

\begin{figure}
    \includegraphics[width=.95\linewidth,keepaspectratio]{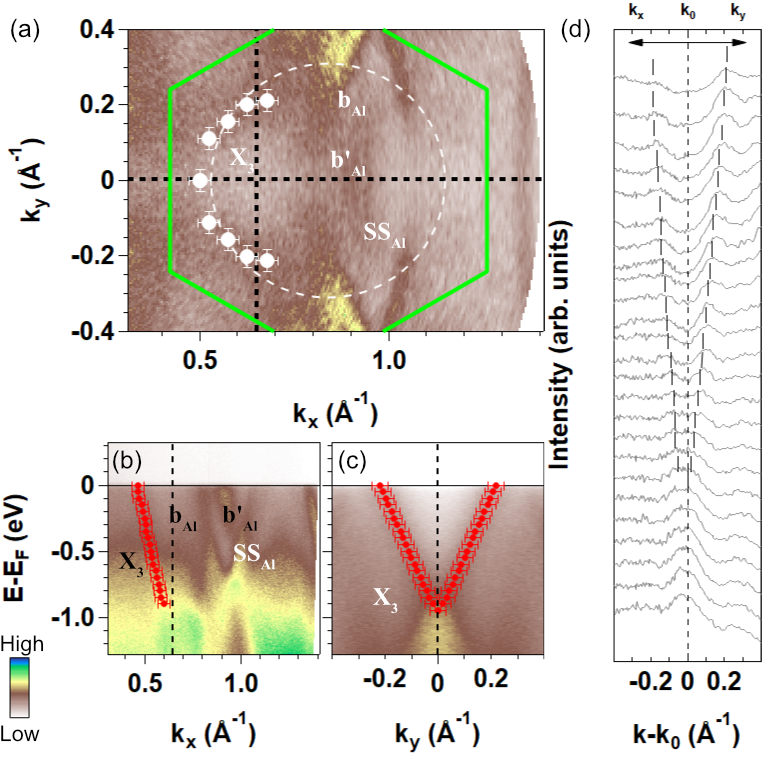}
    \caption{\label{fig4}(a) Fermi surface mapping of 3x3-Si/Al(111) near $\rm M_{1x1}$ point taken at 20 K.
    Green hexagon represents Al(111)-3x3 Brillouin zone. White markers are the positions of Fermi wavevector of $X_{3}$ determined from the peaks in the experimental spectra. White broken line represents the estimated Fermi surface contour made by the $X_{3}$ band.
    (b, c) Band dispersion plot along (b) $k_{y}= \textrm{0 \AA$^{-1}$}$ and (c) $k_{x}=\textrm{0.65 \AA$^{-1}$}$. Red markers are the positions of peaks in momentum distribution curves. (d) Momentum distribution curves near $k_{0}$ = (0.65, 0).  Curves shown in $k-k_{0} < 0$ are along $k_{y}= \textrm{0 \AA$^{-1}$}$ and those in $k-k_{0} > 0$ are along $k_{x}=\textrm{0.65 \AA$^{-1}$}$. Momentum distribution curves are taken for each 0.05 eV slices from E-$\rm E_{F}$ = 0 eV to -1.2 eV. All the error bars correspond to $\pm\textrm{0.03 \AA}^{-1}$.}
\end{figure}

\begin{figure}
    \includegraphics[width=0.9\linewidth,keepaspectratio]{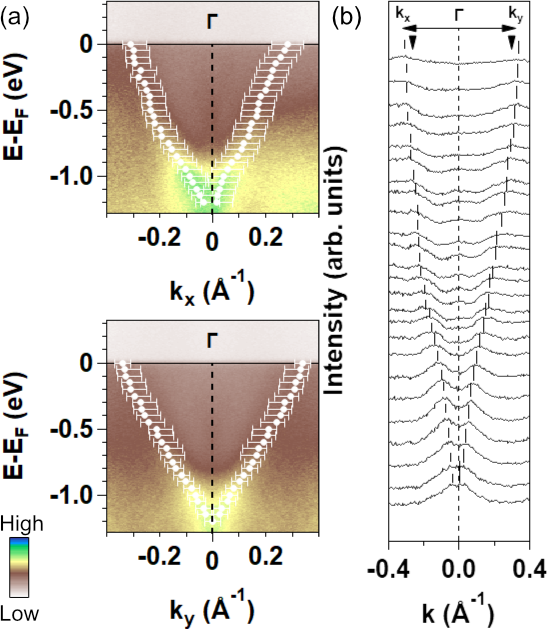}
        \caption{\label{fig3} (a) $\rm k_x$, $\rm k_y$ band-dispersion plots of 3x3-Si/Al(111) at the $\Gamma$ point taken at $h\nu = \textrm{47 eV}$. White markers represent positions of the peaks in experimental spectra.
        (b) Momentum distribution curves near $\Gamma$ point. Curves shown in $k < 0$ are along $k_{y}= \textrm{0 \AA$^{-1}$}$ and those in $k> 0$ are along $k_{x}=\textrm{0 \AA$^{-1}$}$. Momentum distribution curves are taken for each 0.05 eV slices from E-$\rm E_{F}$ = 0 eV to -1.2 eV. Black triangle markers represent the positions where Umklapp replica of $\rm SS_{Al}$ would appear. All the wide and narrow error bars correspond to $\textrm{0.06 \AA}^{-1}$ and $\textrm{0.03 \AA}^{-1}$, respectively.
        }
\end{figure}

\begin{figure}
    \includegraphics[width=.7\linewidth,keepaspectratio]{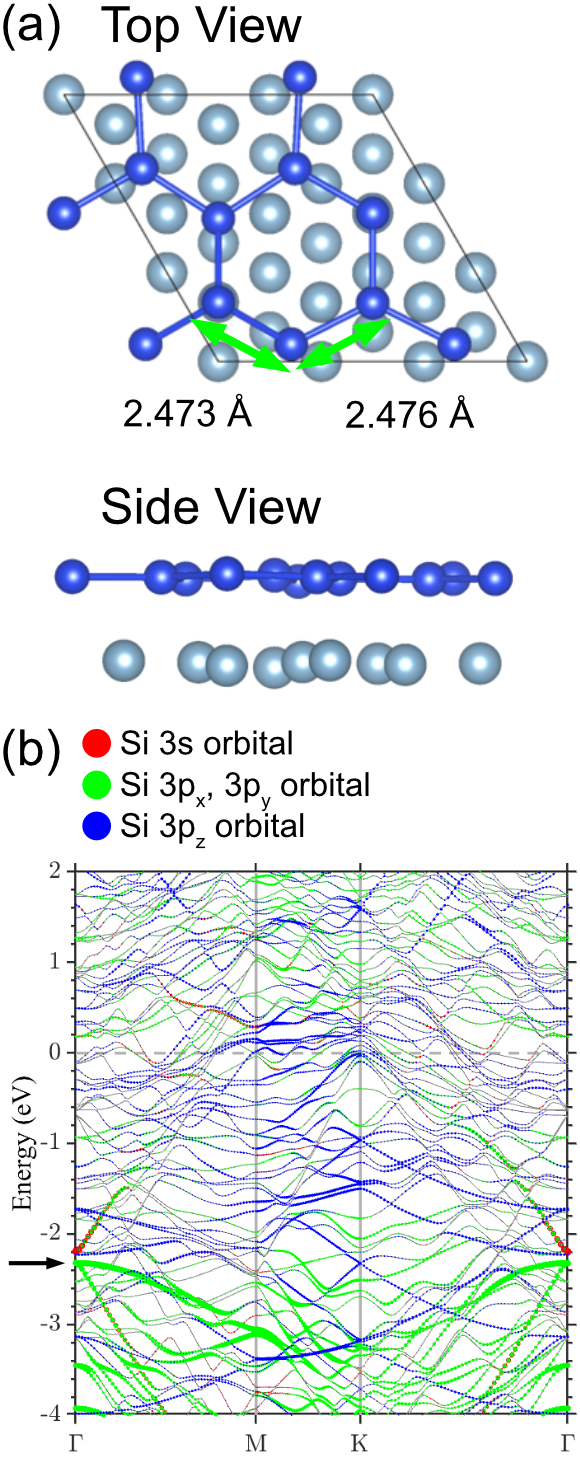}
    \caption{\label{fig6}(a) Schematic drawing of the unit cell of the silicon monolayer with no buckling used in the DFT calculation. (b) Calculated band structure of the silicon layer on an Al(111) substrate. Al(111) slab has six layers. Characters on the bottom axis correspond to the high symmetry points of the Al(111)-3x3 supercell Brillouin zone. The contributions from Si-3$s$, 3$p_{x,y}$, and 3$p_{z}$ orbitals are indicated by
    the size of red, green, and blue dots, respectively. A black arrow indicates the band structure compared to the experimental spectra in Fig. \ref{fig2}.}
\end{figure}
\begin{figure}
    \includegraphics[width=.95\linewidth,keepaspectratio]{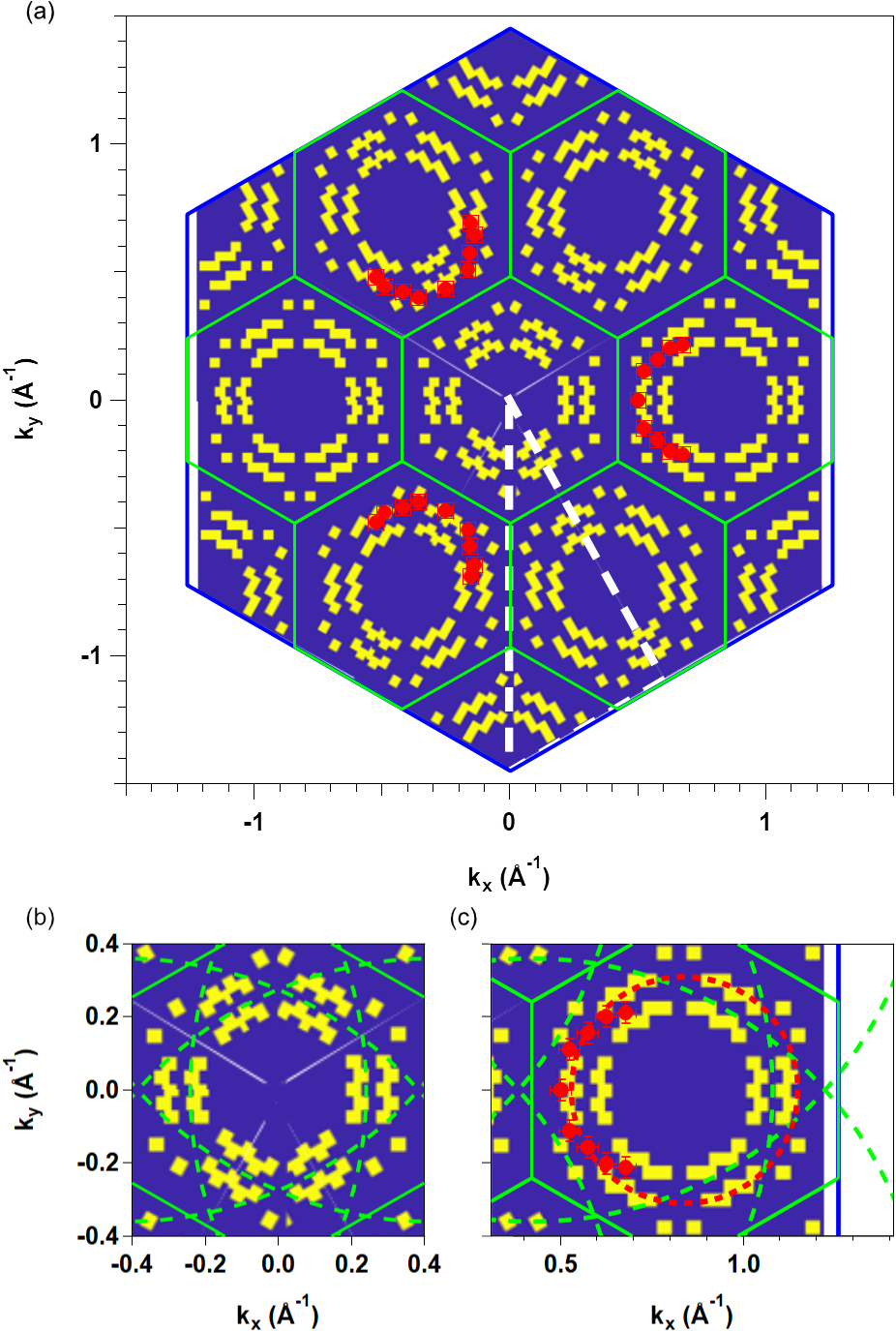}
    \caption{\label{fig7} (a) 
    Fermi surface of the Si/Al(111) constructed by unfolding the band structure of silicene-2x2/Al(111)-3x3 supercell SBZ to Al(111)-1x1 SBZ.
    Yellow dots show the Fermi wavevector. Red markers represent the experimental Fermi wavevector of $X_{3}$. The overall mapping is obtained by the symmetric folding of the region surrounded by white broken curves. (b) Calculated Fermi surface near $\Gamma$ point. Green dashed lines represent the Umklapp replicas of $\rm SS_{Al}$ band in Fig. \ref{fig1}. (c) Calculated Fermi surface near $\rm M_{1x1}$ point. Red markers represent the experimental Fermi wavevector of $X_{3}$. Red dashed line represents the estimated Fermi surface. Green dashed lines represent the Umklapp replicas of $\rm SS_{Al}$ band in Fig. \ref{fig1}. All the error bars correspond to $\pm\textrm{0.03 \AA}^{-1}$.} 
\end{figure}

Focusing on spectral features of the photoemission Fermi surface around $k_{x}=\textrm{0.65 \AA$^{-1}$}$, as shown in the red rectangle in Fig. 1
(d), the Fermi surface map is revisited in Fig. \ref{fig4} (a) with the dispersion plots at (b) $k_{y} = \textrm{0 \AA$^{-1}$}$ and (c) $k_{x}=\textrm{0.65 \AA$^{-1}$}$. The metallic $X_{3}$ band with the linear dispersion and with its crossing point of $\textrm{E-E$_{F}$}=-\textrm{1 eV}$ can be found at $k_{y} = \textrm{0 \AA$^{-1}$}$ and $k_{x}=\textrm{0.65 \AA$^{-1}$}$. 
Possible Fermi surfaces of 3x3-Si/Al(111) are schematically traced in Fig. \ref{fig4}(a), indicating existence of a closed Fermi surface for the $X_{3}$ band.

As shown in Fig. \ref{fig3}, there is another band found at the $\Gamma$ point. That dispersion has uncertainty due to its shoulder structure positioned at {0.03 \AA$^{-1}$} near to $\Gamma$ point. 
In momentum dispersion curves (Fig. \ref{fig3}(b)), the peak positions corresponding 
to this dispersion are near to the position where $\rm SS_{Al}$ could be scattered by the G vector of Al(111)-3x3 SBZ.
In addition, both the dispersion and $\rm SS_{Al}$ show similar band velocity around \SI{e+6}{m/s}. 
From these similarities, the newly observed band is difficult to be distinguished from the 3x3-Umlapp replica of $\rm SS_{Al}$. 

To investigate possible origins of electronic states of 3x3-Si/Al(111), we conducted band calculation on a trial atomic model of the honeycomb Si lattice, a silicene layer, on Al(111). The model structure is illustrated in Fig. \ref{fig6}(a). Figure \ref{fig6} (b) shows the dispersion plots with weights of Si atomic orbitals. At binding energies around -2 eV, dispersion curves of light and heavy effective mass can be identified from $\Gamma$, as $X_{1}$ and $X_{2} (X'_{2})$ in Fig. \ref{fig2}. The agreements between experiments and calculation results imply that the atomic structure may be related to the honeycomb Si lattice and the observed bands originate from 3$p_{x}$ and 3$p_{y}$ orbitals. 

Moving onto electronic states near $\rm E_F$ in Fig. \ref{fig6} (b), states with Si 3$p_{z}$-character can be found and they are examined in terms of the external effect\cite{borophene,Feng2,Liu2006,Mathis}. The band dispersion curves are complicated because of the Si-Al hybridizations and they have the Dirac-cone-like behavior between the symmetry points ($\Gamma$, M). This contradicts a case of the a free-standing silicene layer that has Dirac-cones at the K points. However, our calculated electronic structure is similar to the shape of $X_{3}$ band, which indicates that the state likely originates from the overlayer-substrate interaction. 
Observations of such hybrid states have been reported on borophene/Ag(111)\cite{borophene} and silicene/Ag(111)\cite{Feng2}. Electronic states with $p_{z}$ orbitals in the overlayer are greatly influenced by a substrate as compared with those of $p_{x}$ and $p_{y}$. This is naturally understood that the $p_{z}$ states point to the substrate\cite{borophene,Feng2,Mathis}. Notably Feng $et$ $al$ reported, in the tight-binding regime, that a pair of Dirac-cones in silicene/Ag(111) results from the environmental perturbation to a silicene layer\cite{Feng2}. 

Furthermore, calculated electronic structure in Fig. \ref{fig6}(b) shows metallic bands which is consistent with the ARPES observation (Fig. \ref{fig4}). 
To examine in detail the bands that contain Si orbitals, Figure \ref{fig7}(a) plots the calculated Fermi surfaces with the Fermi wavenumber points depicted by yellow dots,
overlaid by the experimental Fermi wavenumber points of the $X_{3}$ band by red dots. 
The contours of the theoretical Fermi surface are generated by unfolding the bands calculated in the silicene-2x2/Al(111)-3x3 supercell SBZ to the Al(111)-1x1 SBZ and by tiling the triangular region surrounded by white broken lines in Fig. \ref{fig7}(a). One can find in the figure a fairly good matching between the calculation and the experiment, indicating that the $X_{3}$ band may originate from the Si honeycomb layer on Al(111). Figures \ref{fig7} (b) and (c) compare the calculated and experimental contours with those of $\rm Al_{SS}$ bands and the replicas in different SBZ. As shown in Fig. \ref{fig7}(c), the Fermi surface of the $X_{3}$ band is clearly distinguished, while there are almost complete overlaps between the calculated and the Umklapp-scattered bands in the center SBZ (Fig. \ref{fig7}(b)). This again indicates the difficulty in experimentally extracting electronic states of an overlayer on Al(111). 

There are fairly good agreement of the metallic band $X_{3}$ between experiment and calculation. However, dispersion curves of the other bands are quantitatively different from each other. For example, a degenerate band of $X_{1}$ and $X_{2}$ at the $\Gamma$ point is energetically located at -1.5 eV in Fig. \ref{fig2}, while the counterpart in calculation can be found at -2.3 eV in Fig. \ref{fig6}. The Si overlayer on Al(111) has basically the honeycomb structure but the model may require modifications such as structural deformation or addition of Si adatoms. A proper understanding of the electronic states requires band calculation which accounts the appropriate structure model as obtained by diffraction methods\cite{FukayaSi,FukayaGe}.

\section{Conclusion}
We studied the electronic structure of the 3x3 ordered-phase of Si on Al(111) by ARPES using synchrotron radiation and band calculation based on DFT. Electronic states of the Si overlayer are experimentally identified.
A closed Fermi surfaces showing linear dispersing bands like Dirac cone is observed. 
By comparing the ARPES spectra with the calculation result of the honeycomb silicene on Al(111), we determined that these electronic bands originate from the Al-Si hybrid states that may be understood in terms of the substrate perturbation of Dirac cones of a silicene, as reported previously\cite{Feng2}. 

\begin{acknowledgments}
This work was supported by the Grant-in-Aid for Specially Promoted Research (KAKENHI 18H03874) from Japan Society for the Promotion of Science, and by a grant from the Labex MATISSE. The preliminary experiment was performed at facilities of the Synchrotron Radiation Research Organization, the University of Tokyo. The computational work was supported by the Ministry of Science and Technology, Taiwan. H.T. Jeng also thanks the CQT-NTHU-MOE, AS-iMATE-109-13, NCHC and CINC-NTU, Taiwan for technical support. Marie~D'angelo, Polina M.~Sheverdyaeva, and Paolo~Moras are acknowledged for supporting the experiment at Elettra. Baojie Feng and Chi-Cheng Lee are acknowledged for providing information on the silicene systems.

\end{acknowledgments}


\begin{thebibliography}{99}
{
\bibitem{MatsudaTextbook} I. Matsuda ed., \textit{Monatomic Two-Dimensional Layers: Modern Experimental Approaches for Structure, Properties, and Industrial Use} (Elsevier, 2018).  

\bibitem{Neto} A. H. Castro Neto, F. Guinea, N. M. R. Peres, K. S. Novoselov, and A. K. Geim
Rev. Mod. Phys. 81, 109 (2009). 


%%Dirac and Xene
\bibitem{Matthes} L. Matthes et al., J. Phys.: Condens. Matter 25, 395305 (2013).

%%Xene
\bibitem{Lelay}P. Vogt, P. DePadova, C. Quaresima, J. Avila, E. Frantzeskakis, M. C. Asensio, A. Resta, B. Ealet, G. Le Lay, Phys. Rev. Lett. 108, 155501 (2012).
\bibitem{Takamura} A. Fleurence, R. Friedlein, T. Ozaki, H. Kawai, Y. Wang, Y. Yamada-Takamura, Phys. Rev. Lett. 108, 245501 (2012). 
\bibitem{Takagi} C.-L. Lin, R. Arafune, K. Kawahara, N. Tsukahara, E. Minamitani, Y. Kim, N. Takagi, M. Kawai, Apple. Phys. Express 5, 045802 (2012). 
\bibitem{FukayaSi} Y. Fukaya, I. Mochizuki, M. Maekawa, K. Wada, T. Hyodo, I. Matsuda, and A. Kawasuso, Phys. Rev. B 88, 205413 (2013).

\bibitem{Davila} M. E. Davila, L. Xian, S. Cahangirov, A. Rubio, G. LeLay, New J. Physics 16, 095002 (2014).
\bibitem{Tang} C.-H. Lin, A. Huang, W. W. Pai, W.-C. Chen, T.-Y. Chen, T.-R. Chang, R. Yukawa, C.-M. Cheng, C.-Y. Mou, I. Matsuda, T.-C. Chiang, H.-T. Jeng, S.-J. Tang, Phys. Rev. Materials 2, 024003 (2018). 
\bibitem{FukayaGe} Y. Fukaya, I. Matsuda, B. Feng I. Mochizuki, T. Hyodo, and S. Shamoto, 2D Materials 3, 035019 (2016).

%%Photoemission Silicene
\bibitem{Paolo1} S. K. Mahatha, P. Moras, V. Bellini, P. M. Sheverdyaeva, C. Struzzi, L. Petaccia, and C. Carbone, Phys. Rev. B 89, 201416(R) (2014). 

\bibitem{Paolo2}
P. M. Sheverdyaeva, S. K. Mahatha, P. Moras, L. Petaccia and G. Fratesi, G. Onida, C. Carbone, ACS Nano, 11,  975 (2017).

\bibitem{Feng1} B. Feng, H. Li, C.-C. Liu, T.-N. Shao, P. Cheng, Y. Yao, S. Meng, L. Chen, and Kehui Wu, ACS Nano 7, 9049 (2013).
 
\bibitem{Feng2} B. Feng, H. Zhou, Y. Feng, H. Liu, S. He, I. Matsuda, L. Chen, E. F. Schwier, K. Shimada, S. Meng, and K. Wu, Phys. Rev. Lett. 122, 196801 (2019).

\bibitem{SiCal1}
C. Lian and S. Meng, Phys. Rev. B 95, 245409 (2017).

\bibitem{SiCal2}
J.-I. Iwata, Y.-I. Matsushita, H. Nishi, Z.-X. Guo, and A. Oshiyama, Phys. Rev. B 96, 235442 (2017).

%%silicene transistor
\bibitem{Tao} L. Tao, E. Cinquanta, D. Chiappe, C. Grazianetti, M. Fanciulli, M. Dubey, A. Molle and D. Akinwande, Nature Nanotech 10, 227 (2015).

%%Si depo on surfaces, general
\bibitem{Feng2017} B. Feng, B. Fu, S. Kasamatsu, S. Ito, P. Cheng, C.-C. Liu, S. K. Mahatha, P. Sheverdyaeva, P. Moras, M. Arita, O. Sugino, T.-C. Chiang, K. Wu, L. Chen, Y. Yao, and I. Matsuda, Nat. Comm. 8, 1007 (2017).

%%Past 3x3-Si/Al(111)
\bibitem{Jona} F. Jona, J. Appl. Phys. 42, 2557 (1971).
\bibitem{Munoz1} M. C. Munoz, F. Soria, and J. L. Sacedon, Surf. Sci. 189/190, 204 (1987).
\bibitem{Munoz2} M. C. Munoz, F. Soria, and J. L. Sacedon, Surf. Sci. 172, 442 (1986).

%%Al photoemission
\bibitem{Kevan} S. D. Kevan, N. G. Stoffel, and N. V. Smith, Phys. Rev. B 31, 1788 (1985).

%%general


%%Sample

%%calculations
\bibitem{Blochl1994} P. E. Bl\"{o}chl, Phys. Rev. B 50, 17953 (1994).
\bibitem{Kresse1999} G. Kresse and D. Joubert, Phys. Rev. B 59, 1758 (1999).
\bibitem{Kresse1993}G. Kresse and J. Hafner, Phys. Rev. B 47, 558 (1993).
\bibitem{Kresse1994}G. Kresse and J. Hafner, Phys. Rev. B 49, 14251(1994).
\bibitem{Kresse1996} G. Kresse and J. Furthm\"{u}ller, Computational Materials Science 6, 15 (1996).
\bibitem{Kresse1996a} G. Kresse and J. Furthm\"{u}ller, Phys. Rev. B 54, 11169 (1996).

%%Topology
\bibitem{Yan} B. Yan, B. Stadtm\"{u}ller, N. Haag, S. Jakobs, J. Seidel, D. Jungkenn, S. Mathias, M. Cinchetti, M. Aeschlimann, and C. Felser, Nature Communications 6, 10167 (2015).


%%external effect to the overlayers
\bibitem{borophene} B. Feng, O. Sugino, R.-Y. Liu, J. Zhang, R. Yukawa, M. Kawamura, T. Iimori, H. Kim, Y. Hasegawa, H. Li, L. Chen, K. Wu, H. Kumigashira, F. Komori, T.-C. Chiang, S. Meng, and I. Matsuda, Phys. Rev. Lett. 118, 096401 (2017).
\bibitem{Liu2006} C. Liu, I. Matsuda, R. Hobara, and S. Hasegawa, Phys. Rev. Lett. 96, 036803 (2006).
\bibitem{Mathis} M. Cameau, R. Yukawa, C.-H. Chen, A. Huang, S. Ito, R. Ishibiki, K. Horiba, Y. Obata, T. Kondo, H. Kumigashira, H.-T. Jeng, M. D'angelo, and I. Matsuda, Phys. Rev. Materials, 3, 044004 (2019).

%% BANDUP
\bibitem{Medeiros}P. V. C. Medeiros, S. Stafström, and J. Björk, Phys. Rev. B 89, 041407 (2014).

%% substrate
\bibitem{substrate1} M. De Santis, M. Muntwiler, J. Osterwalder, G. Rossi, F. Sirotti,　A. Stuck, and L. Schlapbach, Surf. Sci. 477, 179 (2001).
\bibitem{substrate2} H. -J. Neff, I. Matsuda, M. Hengsberger, F. Baumberger, T. Greber, and J. Osterwalder, Phys. Rev. B 64, 235415 (2001).

}
\end{thebibliography}
\end{document}